\documentstyle[prb,floats,eqsecnum,aps]{revtex}
\begin{document}
\title{\hfill {\bf WUB 94-25} \\[3cm] Static Structure Factors of the 
	XXZ-Model in the presence of a uniform field }

\author{M. Karbach\cite{mailkarbach}, K.-H. M\"utter\cite{mailmuetter} 
	and M. Schmidt\cite{mailschmidt}}
\address{Physics Department, University of Wuppertal\\ 
42097 Wuppertal, Germany}

\date{\today}

\maketitle
\begin{abstract}
The static structure factors of the XXZ model in the presence of uniform field
are determined from an exact computation of the groundstates at given total spin
on rings with $N=4,6,\ldots,28$ sites. In contrast to the naive expectation a
weak uniform field strengthens the antiferromagnetic order in the transverse
structure factor for the isotropic case.

\end{abstract}
\draft
%%%%%%%%%%%%%%%%%%%%%%%%%%%%%%%%%%%%%%%%%%%%%%%%%%%%%%%%%%%%%%%%%%%%%%%%%%%%%%%%
%
%
\section{Introduction}
%
%
%%%%%%%%%%%%%%%%%%%%%%%%%%%%%%%%%%%%%%%%%%%%%%%%%%%%%%%%%%%%%%%%%%%%%%%%%%%%%%%%
The antiferromagnetic properties of the one-dimensional spin $1/2$ XXZ-model 
with Hamiltonian:
\begin{eqnarray}\label{H}
	H=2\sum_{x=1}^N
		\left[S_1(x)S_1(x+1)+S_2(x)S_2(x+1)+\cos \gamma \; S_3(x)S_3(x+1)\right]
\end{eqnarray}
have been studied by analytical \cite{yang} and numerical
\cite{bonner,karbach,fabricius} methods. The critical exponents $\eta_j,j=1,3$,
which govern the large distance behavior of the spin-spin correlators in the
groundstate:
\begin{eqnarray*}
	\langle0|S_j(0)S_j(x)|0\rangle 
	\longrightarrow \frac{(-1)^x}{x^{\eta_j}}\quad j=1,3,
\end{eqnarray*}
are given by:\cite{luther}
\begin{eqnarray*}
	\eta_1=\eta_3^{-1}=1-\frac{\gamma}{\pi}\quad 
	\text{for} \quad 0< \gamma <\pi. 
\end{eqnarray*}
At finite temperature $T$ the spin-spin correlators decrease exponentially with
a rate given by the inverse correlation length:\cite{suzuki}
\begin{eqnarray*}
	\xi^{-1}(\gamma,T)\longrightarrow \Phi(\gamma) \cdot T^{\nu},
\end{eqnarray*}
with a critical exponent:
\begin{eqnarray*}
	\nu=1,
\end{eqnarray*}
independent of $\gamma$ and:
\begin{eqnarray*}
	\Phi(\gamma)=\frac{\gamma}{\sin \gamma}\left(1-\frac{\gamma}{\pi}\right).
\end{eqnarray*}
It has been shown in Ref.\ \onlinecite{karbach} that the structure factors:
\begin{eqnarray*}
	{\bf S}_j(\gamma,p=2\pi k/N,T,N)=1+(-1)^k4\langle0| S_j(0)S_j(N/2)|0\rangle 
	+8\sum_{x=1}^{N/2-1}\langle 0|S_j(0)S_j(x)|0\rangle \cdot \cos(px),
\end{eqnarray*}
are most suited to extract the critical behavior from finite systems. In
Refs.\ \onlinecite{karbach,fabricius} we have studied the static structure
factors in the following three limits:
\begin{eqnarray*}
	p=\pi,\quad T=0,\quad N\rightarrow \infty,
\end{eqnarray*}
\begin{eqnarray*}
p \rightarrow \pi,\quad T=0,\quad N=\infty,
\end{eqnarray*}
\begin{eqnarray*}
p=\pi,\quad T\rightarrow 0,\quad N=\infty.
\end{eqnarray*}
The transverse and longitudinal structure factors have a common form in each of
these limits:
\begin{eqnarray}\label{sjgya}
	{\bf S}_j(\gamma,y_a)=
		r_j(\gamma)\frac{\eta_j(\gamma)}{\eta_j(\gamma)-1}
		\left(1-y_a^{1-\eta_j(\gamma)}\right),\quad a=p,T,N,
\end{eqnarray}
where:
\begin{eqnarray*}
	y_N \equiv \frac{N}{N_j(\gamma)},\quad 
	y_p \equiv \frac{\pi-p_j(\gamma)}{\pi-p},\quad
	y_T \equiv \frac{\xi(\gamma,T)}{\xi_j(\gamma)},
\end{eqnarray*}
are the ``running variables''. The common form reflects the fact that the
structure factors scale in the critical regime:
\begin{eqnarray}\label{ptn}
	p\longrightarrow \pi,\quad T\longrightarrow 0,\quad N\longrightarrow \infty,
\end{eqnarray}
if we keep fixed:
\begin{eqnarray}\label{z1z2}
	z_1 \equiv \left(1-\frac{p}{\pi}\right)N, \quad 
	z_2 \equiv \frac{N}{\xi(\gamma,T)}.
\end{eqnarray}
The antiferromagnetic properties of the model become visible in the
singularities of the structure factors in the limit (\ref{ptn}). The transverse
structure factors are infinite in this limit; they develop a ``hard''
singularity. In contrast, the longitudinal structure factors stay finite. Their
critical behavior is hidden in subleading terms which produce a ``soft''
singularity. Away from the critical regime (\ref{ptn}) the antiferromagnetic
properties are lost.

There are further possibilities to destroy antiferromagnetic ordering -- e.g. by
a uniform external field $h$ or by frustration -- i.e. by switching on a
next-to-nearest neighbor interaction. In this paper we are going to study the
effect of a uniform magnetic field on the zero temperature structure
factors. For this purpose we have determined the static structure factors in the
groundstates $|S_3\rangle$ of the Hamiltonian (\ref{H}) at given total spin
$S_3$. The groundstates were computed with a Lanczos algorithm on rings with
$N=4,6,....,28$ sites.

We will discuss the static structure factors as function of the magnetization
$M=S_3/N$. The known \cite{yang,karbachdis} magnetization curve translates $M$
into the fieldstrength $h$. For convenience, we list in Table I the $h$-field
values corresponding to our $M$-values:
\begin{table}[h]
\caption{magnetization and the corresponding $h$-fields}
\begin{tabular}[h]{c|cccccc}
M&$\slantfrac{1}{8}$&$\slantfrac{1}{6}$& $\slantfrac{1}{4}$&$\slantfrac{1}{3}$&$\slantfrac{3}{8}$&$\slantfrac{1}{2}$\\ 
	\hline
$h(\gamma=0) $     & 0.96 & 1.20 & 1.59 & 1.83 & 1.91 & 2.00 \\
$h(\gamma=0.1\pi)$ & 0.93 & 1.17 & 1.54 & 1.78 & 1.86 & 1.95 \\
$h(\gamma=0.2\pi)$ & 0.83 & 1.06 & 1.41 & 1.64 & 1.72 & 1.81 \\
$h(\gamma=0.5\pi)$ & 0.38 & 0.50 & 0.71 & 0.87 & 0.92 & 1.00 \\
\end{tabular}
\end{table}	

The outline of the paper is as follows. In Sections II and III we discuss the
characteristic features of the longitudinal structure factors as there are the
$p$ and $M$ dependence and the finite-size effects. The same is done for the
transverse structure factors in sections IV and V.
\newpage
%%%%%%%%%%%%%%%%%%%%%%%%%%%%%%%%%%%%%%%%%%%%%%%%%%%%%%%%%%%%%%%%%%%%%%%%%%%%%%%%
%
%
\section{The Longitudinal Structure Factors at Fixed Magnetization}
%
%
%%%%%%%%%%%%%%%%%%%%%%%%%%%%%%%%%%%%%%%%%%%%%%%%%%%%%%%%%%%%%%%%%%%%%%%%%%%%%%%%
The longitudinal structure factor of the XX-model ($\gamma=\pi/2$) is known from
the exact solution obtained by Niemeijer \cite{niemeijer} in the fermion
representation:
\begin{eqnarray}\label{sf3gpid2}
	{\bf S}_3(\gamma=\pi/2,p,M,N) = \frac{2}{\pi} \left\{ \begin{array}{ll}
		 p        & \text{for}  \quad 0 \leq p \leq p_3(M) \\[0.1cm]
		 p_3(M)   & \text{for}  \quad p_3(M)\leq p \leq \pi,
											\end{array} \right.
\end{eqnarray}
where
\begin{eqnarray}\label{p3m}
	p_3(M) \equiv \pi(1-2M).
\end{eqnarray}
Though the result of Ref.\ \onlinecite{niemeijer} has been derived for the
thermodynamical limit $N\rightarrow \infty$, (\ref{sf3gpid2}) turns out to be
correct for all system sizes with $N=4,6,8...$. The linear behavior in $p$ has
been found before \cite{karbach} for the case $M=0$.

Let us next turn to the isotropic case $\gamma=0$. Fig. 1 presents a panoramic
view on the longitudinal structure factor with $N=20,22,...,28$. The emergence
of the singularity at $p=\pi,M=0$ is clearly visible. Along the momentum axis at
$M=0$ we see the logarithmic singularity:
\begin{eqnarray}\label{s3g0m0}
	{\bf S}_3(\gamma =0,p,M=0)=r_3(0) \cdot \ln\left(1-\frac{p}{\pi}\right),
\end{eqnarray}
discussed in Ref.\ \onlinecite{karbach}. Along the magnetization axis:
\begin{eqnarray}\label{pp0}
	p=p_0 \equiv \pi, \quad M=S_3/N\longrightarrow 0,
\end{eqnarray}
we observe a logarithmic singularity in $M$. The same type of singularity is
also found in the limit:
\begin{eqnarray}\label{pp3m}
	p=p_3(M),\quad M \longrightarrow 0,
\end{eqnarray} 
where $p_3(M)$ is given in (\ref{p3m}). The longitudinal structure factor has
its maximum at $p=p_3(M)$, $M$ fixed.  With increasing strength of the uniform
field the maximum position moves from $p=\pi$ to $p=0$ -- i.e. from
antiferromagnetic to ferromagnetic order. Such a behavior has been conjectured
by M\"uller\cite{mueller} {\it et al.}. Indications for this have been found
also by Parkinson and Bonner\cite{parkinson} and Ishimura and
Shiba\cite{ishimura} on small systems ($N\leq 14$). Johnson and
Fowler\cite{johnson} were able to reformulate the isotropic Heisenberg model for
large spins and magnetizations close to saturation in terms of a gas of
magnons. By accident their prediction for the longitudinal structure factor in
the limit $M\rightarrow 1/2$ is identical with the exact result (\ref{sf3gpid2})
for the XX-model. In Fig. 2 we compare the longitudinal structure factors for
$\gamma=0$ and $\gamma=\pi/2$ at $M=1/4$ and $M=1/3$. For $M=1/3$ the structure
factors almost coincide. For smaller $M$ values, however, the isotropic
structure factor deviates from (\ref{sf3gpid2}). The cusp along the line
(\ref{p3m}) becomes more pronounced with decreasing values of $M$.

Looking at the finite-size effects which will be analyzed in the next section we
find an $N^{-2} $ behavior away from the cusp and a slower decrease
$N^{-\delta_3}$ with $\delta_3\approx0.5$ at the cusp $p=p_3(M)$. The change in
the finite-size behavior signals the emergence of a nonanalytic behavior in the
thermodynamical limit.

We have also determined the longitudinal structure factors for the anisotropies
$\gamma/\pi=0.1,0.2$. The $p-M$ dependence of the longitudinal structure factor
looks similar to the isotropic case. Instead of an infinity we find a peak at
$p=\pi,M=0$. In the limit $p\rightarrow \pi,M=0$ the structure factor is
adequately described by (\ref{sjgya}) with $a=p$. In the limit $M\rightarrow
0,p=\pi$ we find again a behavior of the form (\ref{sjgya}) with a running
variable:
\begin{eqnarray}\label{ym}
	y_M=\frac{M}{M(\gamma)}.
\end{eqnarray}
The appearence of the cusp along the line $p=p_3(M)$ is indeed 
independent of the anisotropy parameter $\gamma$. For increasing
values of $\gamma$ the cusp is less pronounced.
%%%%%%%%%%%%%%%%%%%%%%%%%%%%%%%%%%%%%%%%%%%%%%%%%%%%%%%%%%%%%%%%%%%%%%%%%%%%%%%%
%
%
\section{Finite Size Analysis of the Longitudinal Structure Factor}  
%
%
%%%%%%%%%%%%%%%%%%%%%%%%%%%%%%%%%%%%%%%%%%%%%%%%%%%%%%%%%%%%%%%%%%%%%%%%%%%%%%%%
In the critical regime
\begin{eqnarray}\label{limitpNmT}
	p \longrightarrow \pi, \quad N \longrightarrow \infty, \quad
	M = \frac{S_3}{N} \longrightarrow 0, \quad T \longrightarrow 0, 
\end{eqnarray}
we expect the longitudinal structure factors to
obey finite-size scaling:
\begin{eqnarray}\label{s3g3}
	{\bf S}_3(\gamma,p,M=S_3/N,T,N)=g_3(\gamma;z_1,z_2,z_3) \cdot
	{\bf S}_3(\gamma,p,M,T,N=\infty)
\end{eqnarray}
In the combined limit (\ref{limitpNmT}) we have to keep fixed $z_1,z_2$ --
defined in (\ref{z1z2})-- and:
\begin{eqnarray*}
	z_3 \equiv M \cdot N=S_3.
\end{eqnarray*}
In Ref.\ \onlinecite{fabricius} we checked finite-size scaling in the combined
limit:
\begin{eqnarray*}
	T \longrightarrow 0,\quad N \longrightarrow \infty, 
	\quad z_2 \quad \text{fixed and }\quad p=\pi, \quad M=0.
\end{eqnarray*}
It was found that finite-size scaling works for the transverse structure factors
at all $\gamma $ values. In contrast, finite-size scaling breaks down for the
longitudinal structure factors, if $\gamma>0.3\pi$.

In this section we are going to study consequences of the ansatz (\ref{s3g3}) in
the limits (\ref{pp0}) and (\ref{pp3m}) at zero-temperature. The behavior of the
isotropic structure factor can be read of (\ref{sjgya}):
\begin{eqnarray*}
	{\bf S}_3(\gamma=0,p=\pi,M,N=\infty) \stackrel{M\to 0}{\longrightarrow}
		r_3(0) \cdot \ln \frac{M}{M_i},
\end{eqnarray*}
where $i=0,3$ stands for the two limits (\ref{pp0},\ref{pp3m}).  The isotropic
structure factors on finite systems with $N$ sites scale in $M$ (for $M\neq 0$)
and are linear in $-\ln M$ in the limit $M\rightarrow 0$, as can be seen from
Fig. 3. Finite-size effects are small for $p_0=\pi$ but rather large for
$p_3=\pi(1-2M)$. We have studied the finite-size effects in the differences:
\begin{eqnarray}\label{DjMN}
	\Delta_j(\gamma,M,N)={\bf S}_3(\gamma=0,p_j,M,N) + \ln (2M),\quad j=0,3,
\end{eqnarray}
at fixed magnetizations:
\begin{eqnarray*}
	M = \left\{ \begin{array}{ll}
		\frac{1}{8}, \frac{3}{8} & : \quad N=8,16,24 \\[0.1cm] 
		\frac{1}{4}				 & : \quad N=4,8,12,16,20,24,28 \\[0.1cm] 
		\frac{1}{6}, \frac{2}{6} & : \quad N=6,12,18,24,
		\end{array} \right.
\end{eqnarray*}
which can be realized on the systems with size $N$. The $N$-dependence of the
difference (\ref{DjMN}) can be parametrized by:
\begin{eqnarray}\label{D0D3}
	\Delta_0(\gamma,M,N)&=&\Delta_0(\gamma,M)+c_0(\gamma,M)\cdot N^{-2}, \\
	\Delta_3(\gamma,M,N)&=&\Delta_3(\gamma,M)+c_3(\gamma,M)\cdot N^{-\delta_3}, 
		\quad \delta_3 \approx 0.5.
		 \nonumber
\end{eqnarray}
In our finite-size analysis we have included as well the magnetizations:
\begin{eqnarray*}
	M = \left\{ \begin{array}{ll}
	\frac{1}{10},\frac{2}{10},\frac{3}{10},\frac{4}{10}&:\quad N=10,20\\[0.1cm] 
	\frac{1}{12},\frac{5}{12}						   &:\quad N=12,24\\[0.1cm] 
	\frac{1}{14},\frac{2}{14},\frac{3}{14},
			\frac{4}{14},\frac{5}{14},\frac{6}{14}	   &:\quad N=14,28,
				\end{array}\right.
\end{eqnarray*}
which occur on two systems. Here we have assumed that the finite-size dependence
is described correctly by (\ref{D0D3}). The extrapolations of the structure
factors to the thermodynamical limit are represented by the the solid dots in
Fig. 4.

So far our estimates of the thermodynamical limit are restricted to
magnetizations $M\geq 1/14$ due to the finiteness of our systems $N\leq
28$. Additional information on the structure factors for $M$-values:
\begin{eqnarray}\label{MdN}
	M=\frac{1}{N}, \quad N=14,16,...,28,
\end{eqnarray}
closer to the critical point $M=0$ can be obtained if finite-size scaling
(\ref{s3g3}) holds for the longitudinal structure factors in the limits
(\ref{pp0},\ref{pp3m}) where we keep fixed $z_3=S_3$.  This variable is just $1$
for the sequence (\ref{MdN}), and we can compute the scaling function at this
value from:
\begin{eqnarray*}
	g_j(\gamma,z_3)\Big|_{z_3=1}=
		\frac{{\bf S}_3(\gamma=0,p=p_j,M=1/14,N=14)}
		{{\bf S_3}(\gamma=0,p=p_j,M=1/14,N=\infty)}.
\end{eqnarray*}
In this way we get from the finite-size scaling ansatz (\ref{limitpNmT}) in the
limits (\ref{pp0},\ref{pp3m}) an estimate of the thermodynamical limit of the
structure factors:
\begin{eqnarray}\label{s3inf}
	{\bf S}_3(\gamma=0,p=p_j,M,N=\infty) =
	\frac{{\bf S}_3(\gamma=0,p=p_j,M=1/N,N)}{g_j(\gamma,1)}, \quad j=0,3,
\end{eqnarray}
for the sequence of $M$ values in (\ref{MdN}). The result for the differences
(\ref{D0D3}) is marked by the open dots in Figs. 4(a),(b).

We have repeated the finite-size analysis -- described above for the isotropic
case -- for: $ \gamma/\pi=0.1,0.2$. The resulting estimate of the
thermodynamical limit:
\begin{eqnarray*}
	\Delta_j(\gamma,M,N=\infty)={\bf S}_3(\gamma,p_j,M,N=\infty)-L_3(\gamma,M),
		\quad j=0,3,
\end{eqnarray*}
versus the variable:
\begin{eqnarray*}
	L_3(\gamma,M)=\frac{\eta_3(\gamma)}{\eta_3(\gamma)-1}
			\left(1-(2M)^{1-\eta_3(\gamma)}\right),
\end{eqnarray*}
is represented in Figs. 4(a),(b) by the triangles and squares, respectively.
%%%%%%%%%%%%%%%%%%%%%%%%%%%%%%%%%%%%%%%%%%%%%%%%%%%%%%%%%%%%%%%%%%%%%%%%%%%%%%%%
%
%
\section{The Transverse Structure Factors at Fixed Magnetization}
%
%
%%%%%%%%%%%%%%%%%%%%%%%%%%%%%%%%%%%%%%%%%%%%%%%%%%%%%%%%%%%%%%%%%%%%%%%%%%%%%%%%
The most remarkable property of the transverse structure factor is its
approximate constancy:
\begin{eqnarray}\label{s1gpmt}
	{\bf S}_1(\gamma,p,M,T=0,N) \approx 2M, \quad \text{for}
	 \quad 0\leq p\leq 2\pi\,M,
\end{eqnarray}
which has been found by M\"uller\cite{mueller} {\it et al.} on small systems for
the isotropic case $\gamma=0$. For $S_3=MN=N/2-1$ (\ref{s1gpmt}) can be easily
proven to be exact by means of the Bethe ansatz solution. For $S_3<N/2-1$ and
$0<p\leq 2M\pi$, however, (\ref{s1gpmt}) is not exact. As an example we present
in Figs. 5(a),(b) the momentum dependence of ${\bf S}_1$ at $M=1/4,1/3,~
N=8,12,...,28$ for $\gamma=\pi/2$ and $\gamma=0$, respectively. The constancy
in the regime $p\leq 2\pi M$ is striking. Deviations from (\ref{s1gpmt}) can be
seen on a scale magnified by a factor 100 in the inset of Figs. 5(a),(b). These
deviations follow a single scaling curve which increases monotonically with
momentum $p$. On the scaling curve finite-size effects die out with $N^{-2}$. At
$p=2M\pi$, however, significant finite-size effects of the order $N^{-\delta_1}$
with $\delta_1\approx 1.0$ for $\gamma=0 $ and $\delta_1\approx 1.3$ for
$\gamma=\pi/2$ become apparent.

Beyond the regime (\ref{s1gpmt}) the transverse structure factor of the XX-model
($\gamma=\pi/2$) is linear in $(1-p/\pi)^{-1/2}$, as can be seen from
Fig. 5(a). Therefore we find the same type of singularity for $p\rightarrow \pi$
at $M=0,1/4,1/3$. In contrast to the naive expectation antiferromagnetic order
is not destroyed in the transverse structure factors by an external field.

In the isotropic case ($\gamma=0$) the transverse structure factor is
approximately linear in $-\ln (1-p/\pi)$ for $p>2M\pi$, as can be seen from
Fig. 5(b). This type of singularity was found for $p\rightarrow \pi$ at
$M=0$. The slight curvature at $M=1/4$ might indicate that the type of the
singularity has changed here to a power behavior $(1-p/\pi)^{-\alpha}$.

In the anisotropic case with $\gamma/\pi = 0.1,0.2$ and $M=1/4$ we find again
the constant behavior (\ref{s1gpmt}) and a linear increase in
$(1-p/\pi)^{\eta_1-1}$. Again this type of singularity follows from
(\ref{sjgya}) with $a=p$ and $M=0$.
%%%%%%%%%%%%%%%%%%%%%%%%%%%%%%%%%%%%%%%%%%%%%%%%%%%%%%%%%%%%%%%%%%%%%%%%%%%%%%%%
%
%
\section{The Transverse Structure Factor at Critical Momentum}
%
%
%%%%%%%%%%%%%%%%%%%%%%%%%%%%%%%%%%%%%%%%%%%%%%%%%%%%%%%%%%%%%%%%%%%%%%%%%%%%%%%%
Let us start with the XX-model ($\gamma=\pi/2$). Fig. 6(a) shows the transverse
structure factor for $p=\pi$ and $N=8,10,...,28$ as function of $M$.  In
contrast to the longitudinal case we have no scaling in $M$. At $M=0$ we know
from (\ref{sjgya}) with $a=N$ that the transverse structure factors diverge as
$\sqrt{N}$.  The same type of divergence appears at $M=1/4$, as can be seen from
Fig. 7(a). Here we compare the $N$-dependence of ${\bf
S}_1(\gamma=\pi/2,p=\pi,M,N)$ for $M=0$ and $M=1/4$. In both cases there is a
linear increase in $\sqrt{N}$. 

Let us now turn to the isotropic case ($\gamma=0$). Here the longitudinal and
transverse structure factors coincide provided that there is no external
field. In the presence of a uniform field, however, they differ drastically.  At
$p=\pi$, the longitudinal structure factor is infinite at $M=0$, but becomes
finite and monotonically decreasing for $M>0$. In contrast the transverse
structure factor stays at infinity for $M>0$. Moreover, on finite systems, the
longitudinal structure factor scales for $M>0$ -- as was demonstrated in Fig. 3
-- whereas the corresponding transverse structure factor (at $p=\pi$) does not
scale at all, as can be seen from Fig. 6(b). More surprising, at fixed $M>0$ not
too large, the transverse structure factors increase with the system size $N$
stronger than at $M=0$. A comparison of the $N$-dependence at $M=0$ and $M=1/4$
is shown in Fig. 7(b).  At $M=1/4$ the increase with $\ln N$ is definitely
steeper, which signals a strengthening of the singularity at $p=\pi$. Note also
that there are deviations from linearity in $\ln N$, which increase with
$M$. This might indicate a change from a logarithmic behavior at $M=0$ to a
power behavior for $M>0$.

In Ref.\ \onlinecite{mueller} M\"uller {\it et al.} reported on the transverse
structure factor for $N=10$; they found already that the dominant mode remains
situated at $p=\pi$ independent of the field. Fig. 6(b) tells us that the
strengthening of the singularity at $p=\pi$ becomes more and more pronounced
with increasing system size $N$.
%%%%%%%%%%%%%%%%%%%%%%%%%%%%%%%%%%%%%%%%%%%%%%%%%%%%%%%%%%%%%%%%%%%%%%%%%%%%%%%%
%
%
\section{Conclusions}
%
%
%%%%%%%%%%%%%%%%%%%%%%%%%%%%%%%%%%%%%%%%%%%%%%%%%%%%%%%%%%%%%%%%%%%%%%%%%%%%%%%%
In the presence of a uniform external field in $z$-direction the static
structure factors of the XXZ-model show up the following features:
\begin{enumerate}
\item
The longitudinal structure factors have a cusp along the line (\ref{p3m}).  In
case of the XX-model ($\gamma=\pi/2$) there are no finite-size effects and the
longitudinal structure factor is given by (\ref{sf3gpid2}) for all system sizes
with $N=4,6,...$. For smaller values of $\gamma$ and $M$, the cusp becomes
sharper. Finite-size effects decrease along the cusp with $N^{-\delta_3}$,
$\delta_3\approx 0.5$. Away from the cusp we find a more rapid decrease of the
order $N^{-2}$.
\item
	The longitudinal structure factor is finite for $\gamma\neq 0$ and
	$p\rightarrow \pi,M\rightarrow 0$, but develops a logarithmic singularity in
	this limit for the isotropic case ($\gamma=0$). This means: A uniform field
	{\bf weakens} the antiferromagnetic order in the {\bf longitudinal}
	structure factor for $\gamma=0$.
\item
	The transverse structure factor is almost contant for $p\leq 2M\pi$
	[cf.(\ref{s1gpmt})]. Finite-size effects die out slowly with $N^{-\delta_1}$
	with $\delta_1\approx 1$ along the line $p=2M\pi$, but rapidly with $N^{-2}$
	away from this line.
\item
	In the limit $p\rightarrow \pi, M=1/4$ fixed, we observe a singularity of
	the type $(1-p/\pi)^{-1/2}$ in the transverse structure factor for
	$\gamma=\pi/2$. In the isotropic case ($\gamma=0$) this singularity appears
	to be stronger than $-\ln (1-p/\pi)$.  This means: A weak uniform field {\bf
	strengthens} the antiferromagnetic order in the {\bf transverse} strucure
	factor for $\gamma=0$.
\end{enumerate}
Therefore the effect of a uniform external field on the longitudinal and
transverse structure factors for ($\gamma=0$ ) is similar to the effect of
switching on the anisotropy parameter $\gamma$. The logarithmic singularity
found in the isotropic structure factor at $p=\pi$ changes with $\gamma$. It is
strengthened in the transverse, but weakened in the longitudinal structure
factor.
\newpage
%%%%%%%%%%%%%%%%%%%%%%%%%%%%%%%%%%%%%%%%%%%%%%%%%%%%%%%%%%%%%%%%%%%%%%%%%%%%%%%%
%
%
{\large \bf FIGURE CAPTIONS}
%
%
%%%%%%%%%%%%%%%%%%%%%%%%%%%%%%%%%%%%%%%%%%%%%%%%%%%%%%%%%%%%%%%%%%%%%%%%%%%%%%%%
\begin{itemize}
\item[FIG. 1~~~~]
	The longitudinal structure factor versus momentum $p$ and magnetization $M$ 
	with a ridge along the line $p_3(M) = \pi(1-2M)$, for $N=20,22,...,28$.
\item[FIG. 2~~~~]
	Comparison of the longitudinal structure factors at $\gamma=\pi/2$ [Eq.
	(\ref{sf3gpid2})] and $\gamma=0$ for $M=1/4\; (\circ), 1/3\;(\bullet)$,
	respectively.
\item[FIG. 3~~~~]
	Scaling of the longitudinal structure factors ${\bf S}_3(\gamma=0,p,M,N),~
	j=0,3$ with $p_0=\pi$ (open symbols), $p_3=\pi(1-2M)$ (solid symbols), 
	with the magnetization $M$.
\item[FIG. 4(a)]
	Estimate of the thermodynamical limit for the difference
	$\Delta_0(\gamma,M,N=\infty)$ [Eq. (\ref{DjMN})] at momentum $p_0=\pi$:
	Solid symbols represent results from the finite-size analysis (\ref{D0D3}),
	open symbols are results from finite-size scaling (\ref{s3inf}).
\item[FIG. 4(b)]
	Same as Fig. 4(a) for the difference $\Delta_3(\gamma,M,N=\infty)$ at the
	cusp $p_3(M) = \pi(1-2M)$.
\item[FIG. 5(a)]
	The momentum dependence of the transverse structure factor of the XX model
	$(\gamma=\pi/2)$ at fixed magnetizations $M=1/4\; (\circ),
	M=1/3\;(\bullet)$, respectively. The inset shows a magnification of the low
	momentum regime.
\item[FIG. 5(b)]
	Same as Fig. 5(a) for the transverse structure factor in the isotropic case
	$(\gamma=0)$. The inset shows a magnification of the low momentum regime.
\item[FIG. 6(a)]
	The transverse structure factor of the XX model $(\gamma=\pi/2)$ versus
	magnetization $M$ at momentum $p=\pi$.
\item[FIG. 6(b)]
	Same as Fig. 6(a) for the isotropic model $(\gamma=0)$.
\item[FIG. 7(a)]
	Comparison of the size dependence of the transverse structure factors of the
	XX model $(\gamma=\pi/2)$ at $p=\pi$: $M=1/4\; (\circ), M=1/3\;(\bullet)$,
	respectively.
\item[FIG. 7(b)]
	Same as Fig. 7(a) for the isotropic model $(\gamma=0)$.
\end{itemize}
\newpage
%%%%%%%%%%%%%%%%%%%%%%%%%%%%%%%%%%%%%%%%%%%%%%%%%%%%%%%%%%%%%%%%%%%%%%%%%%%%%%%%
%
%

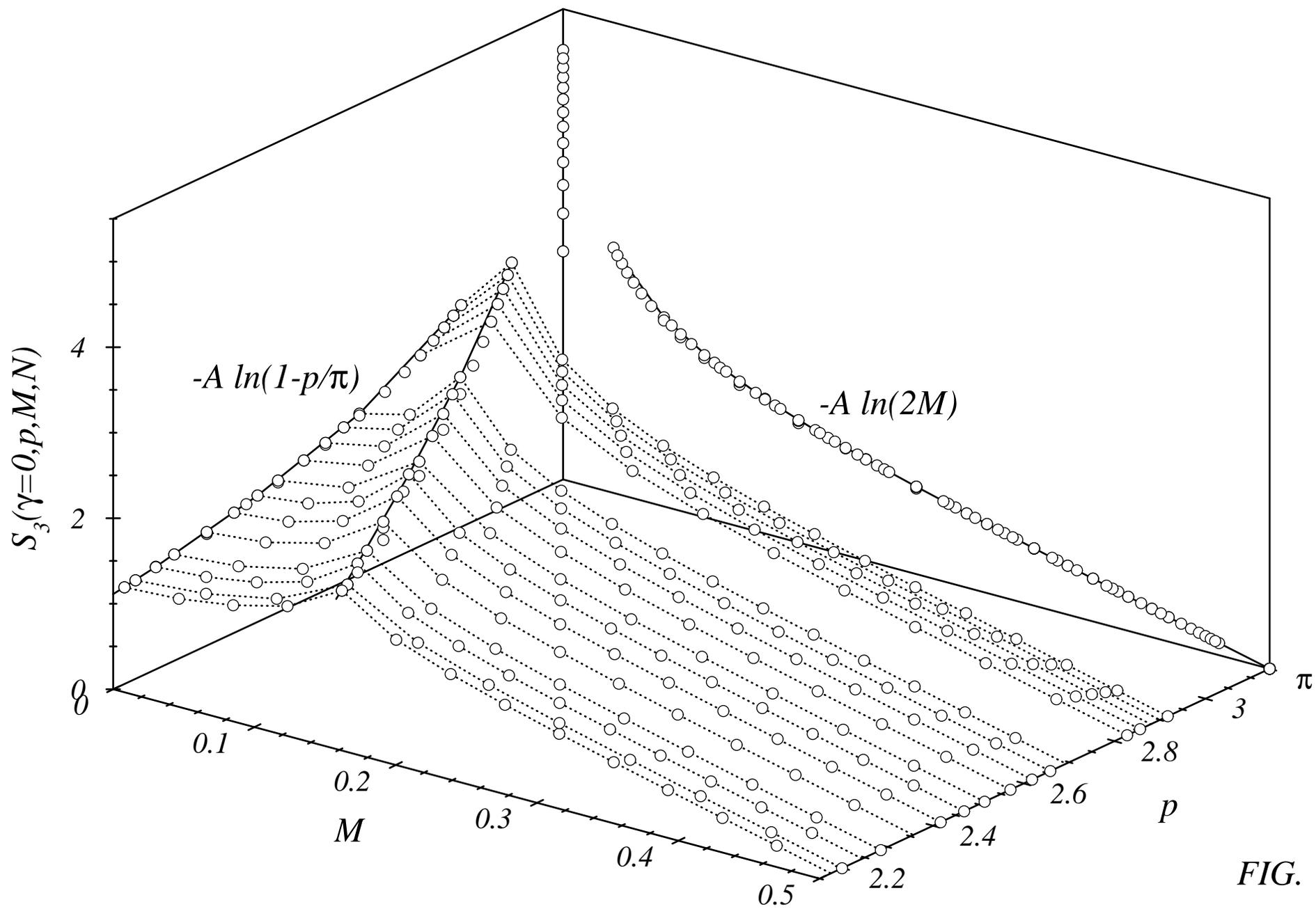
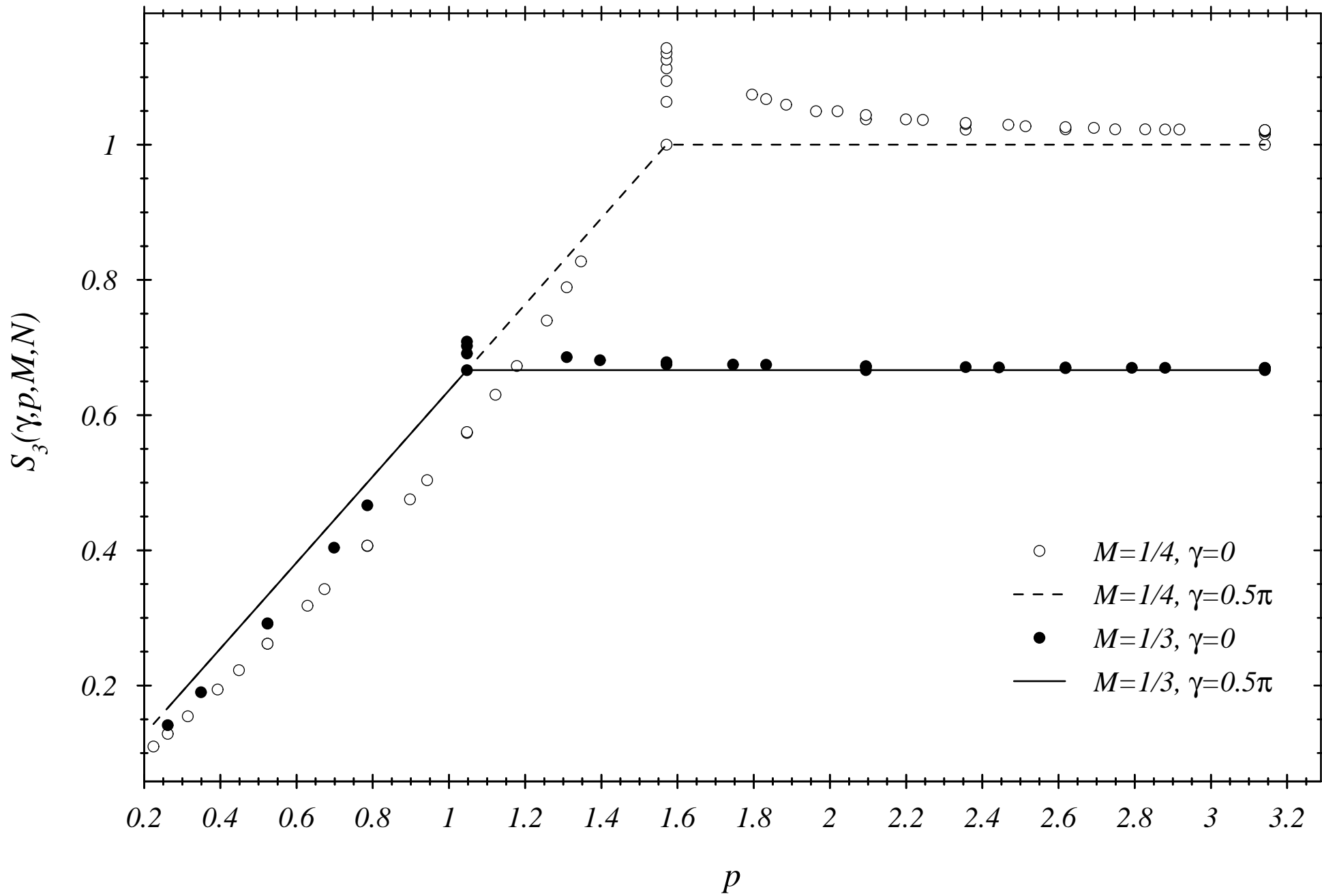
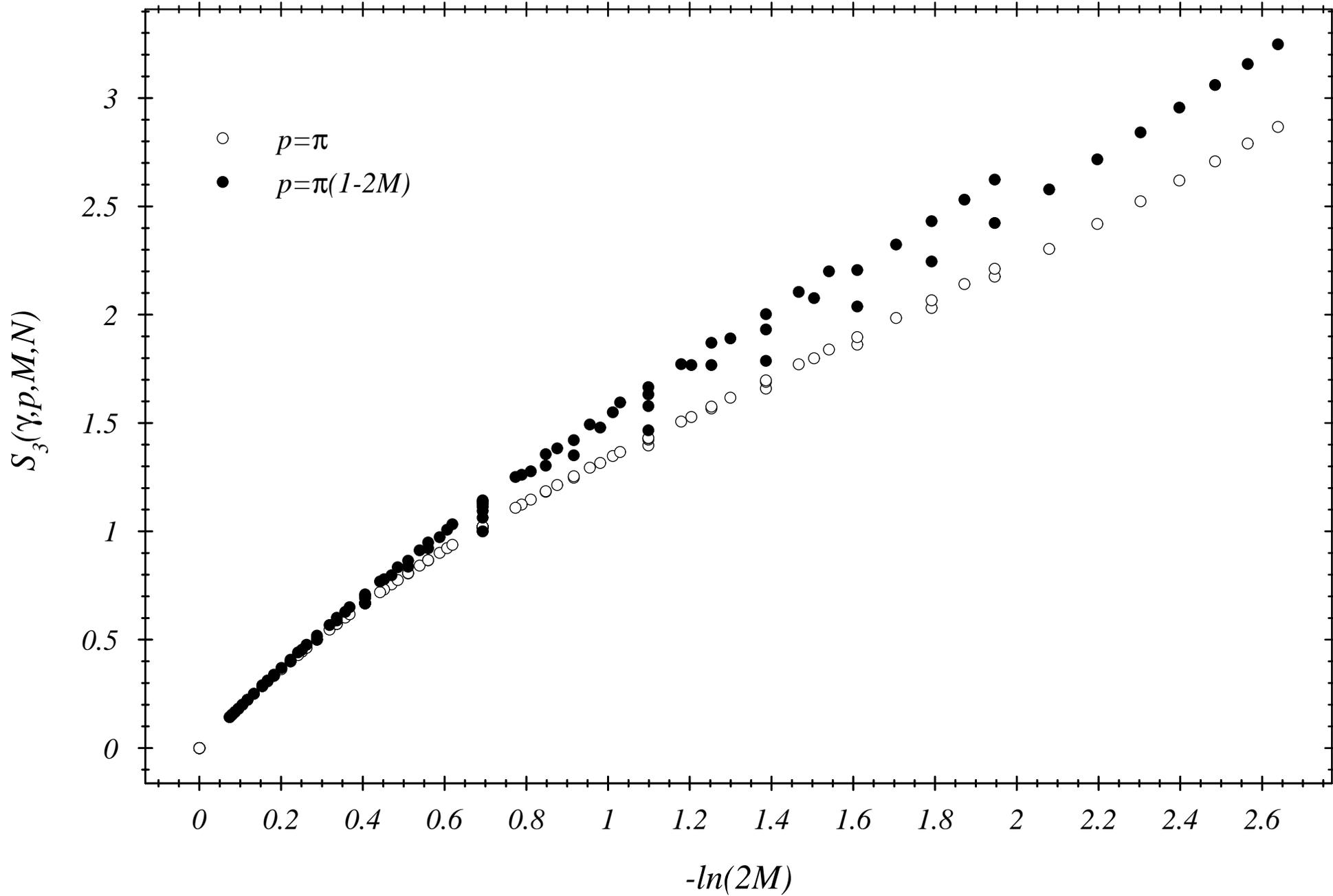
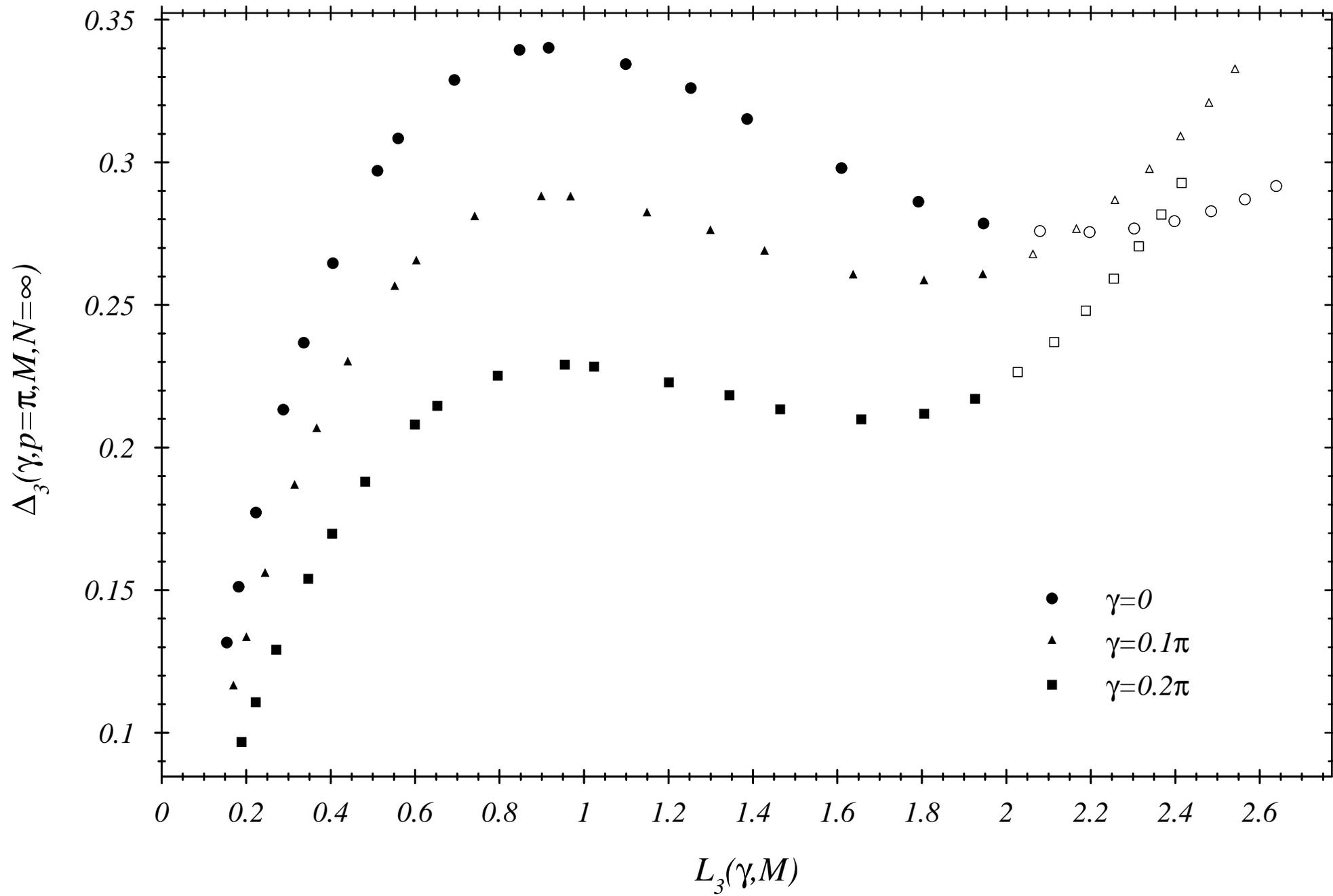
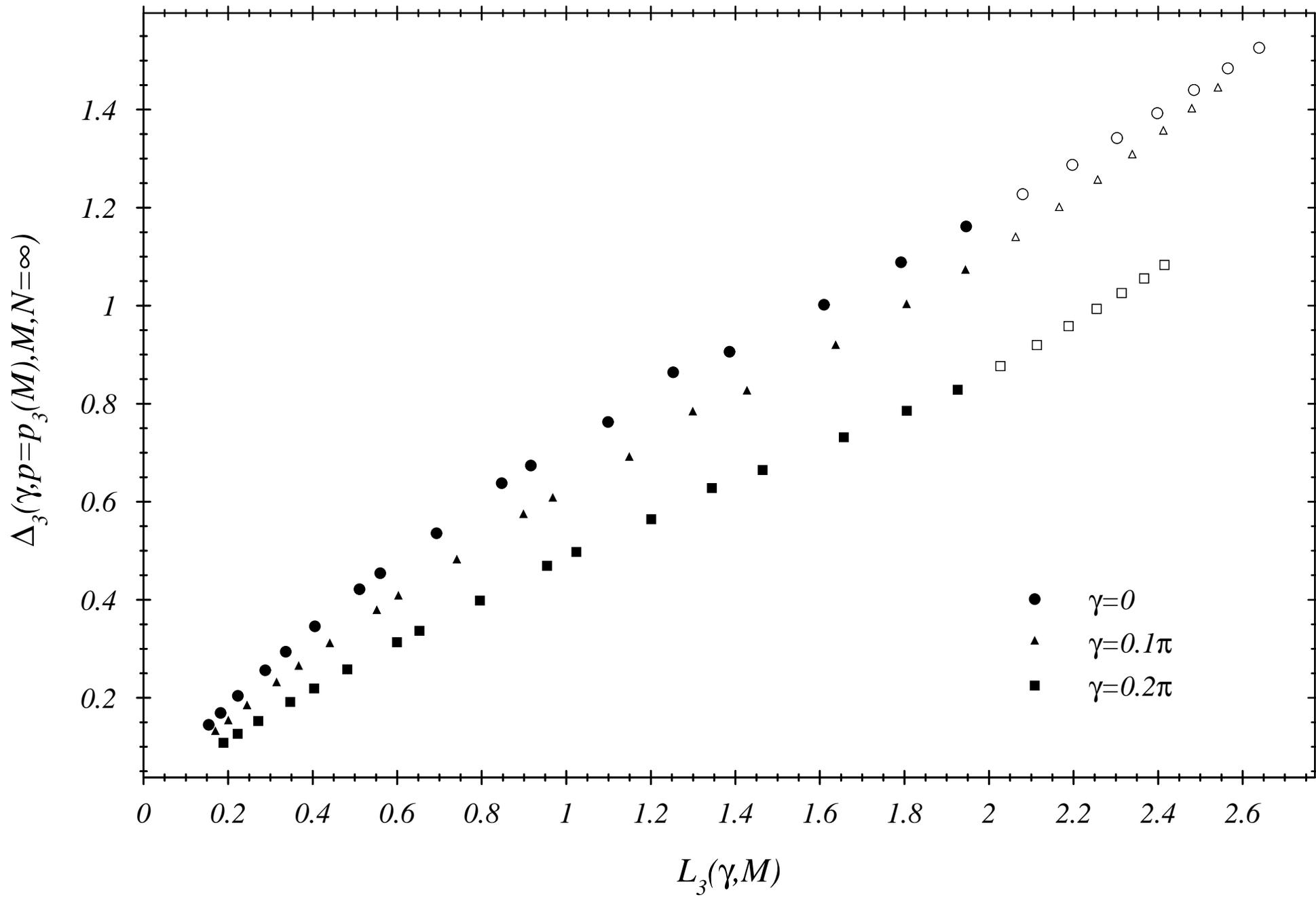
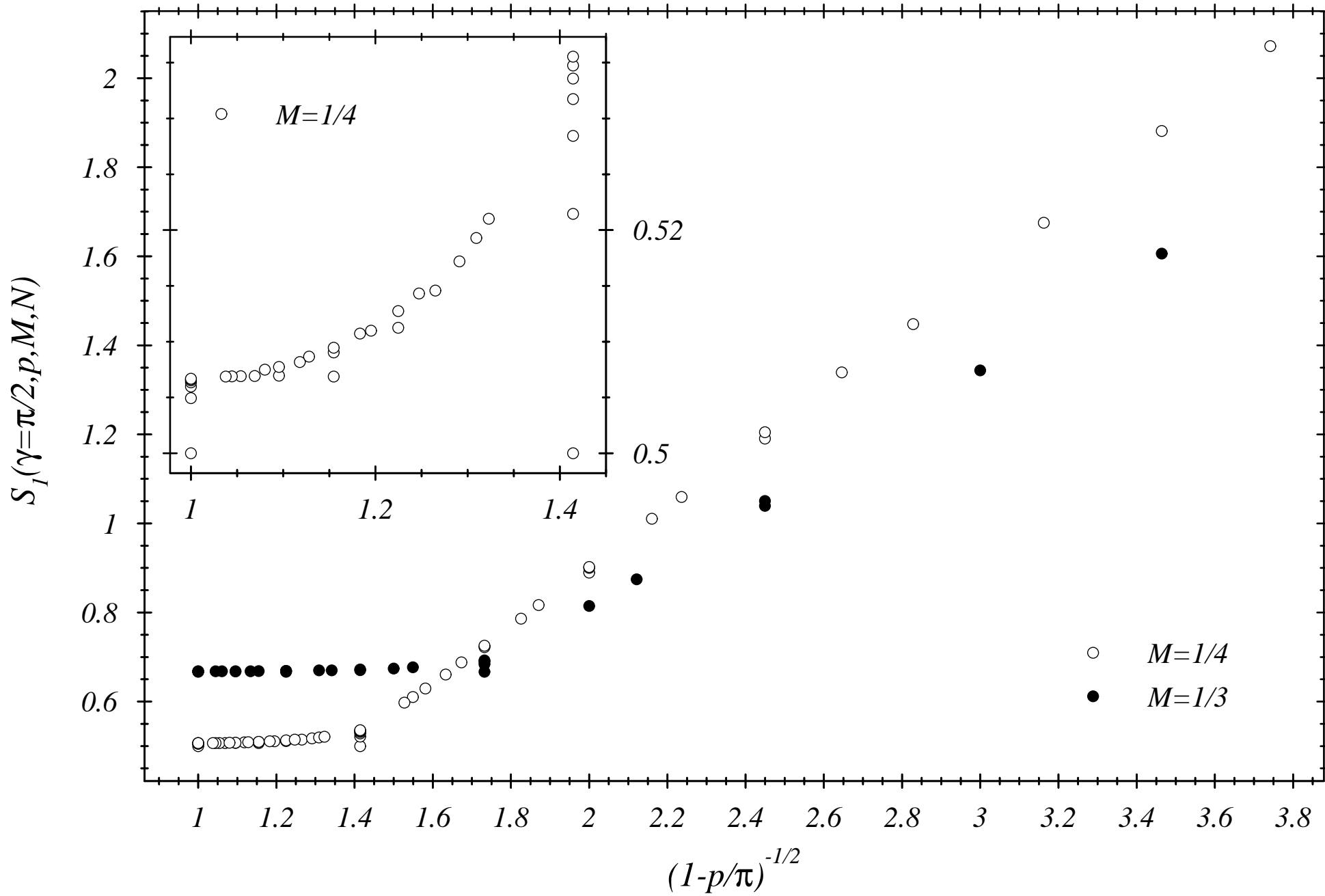
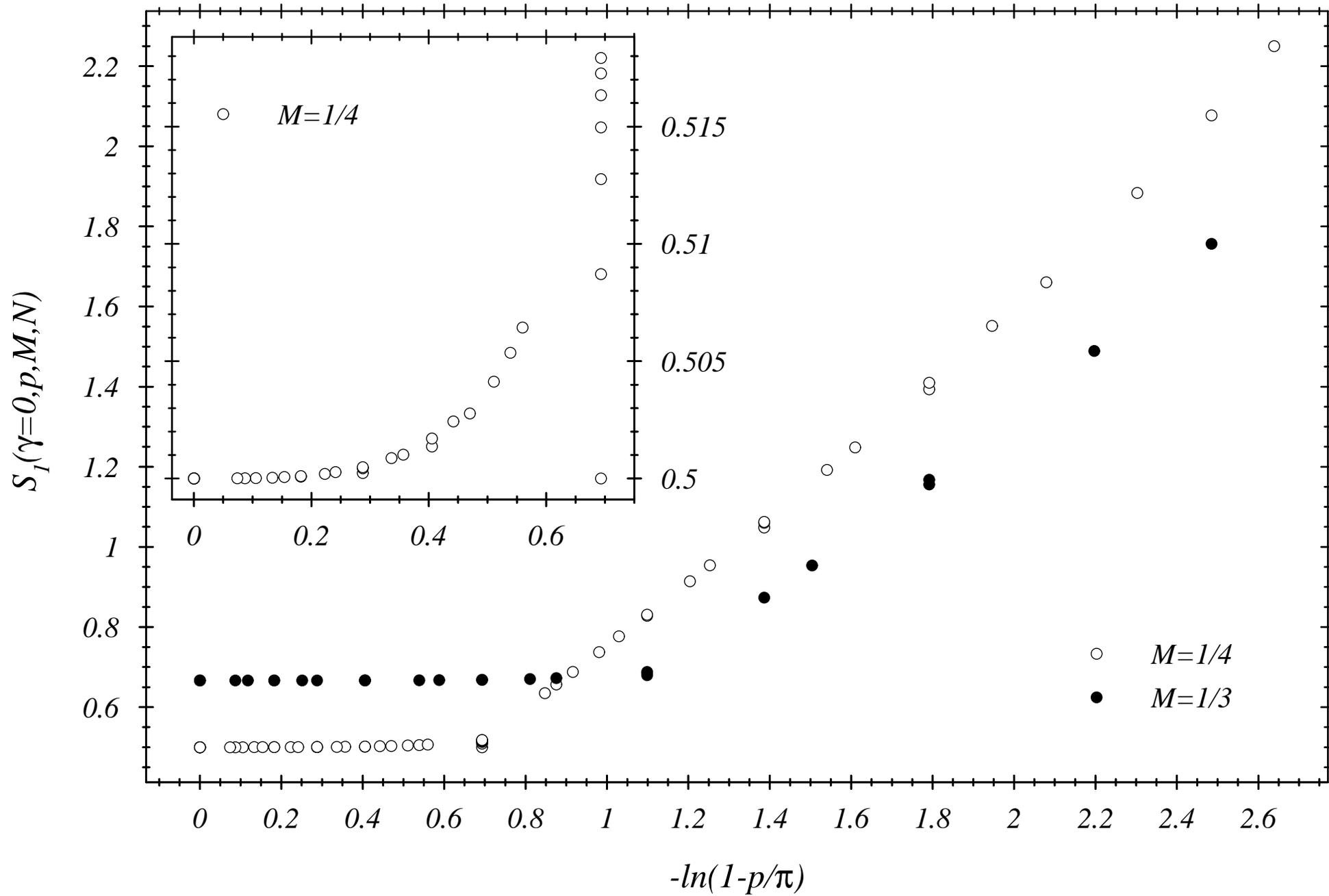
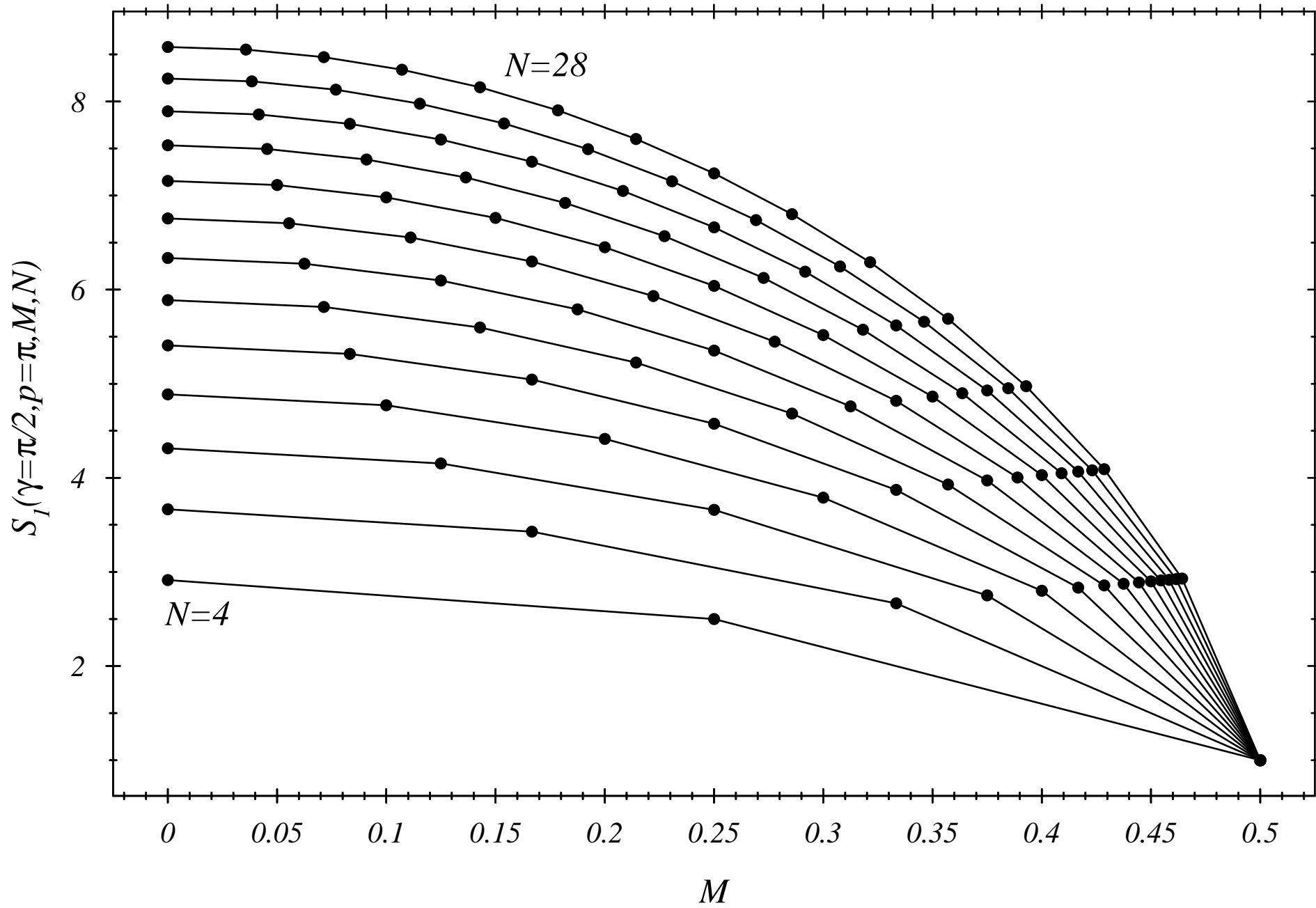
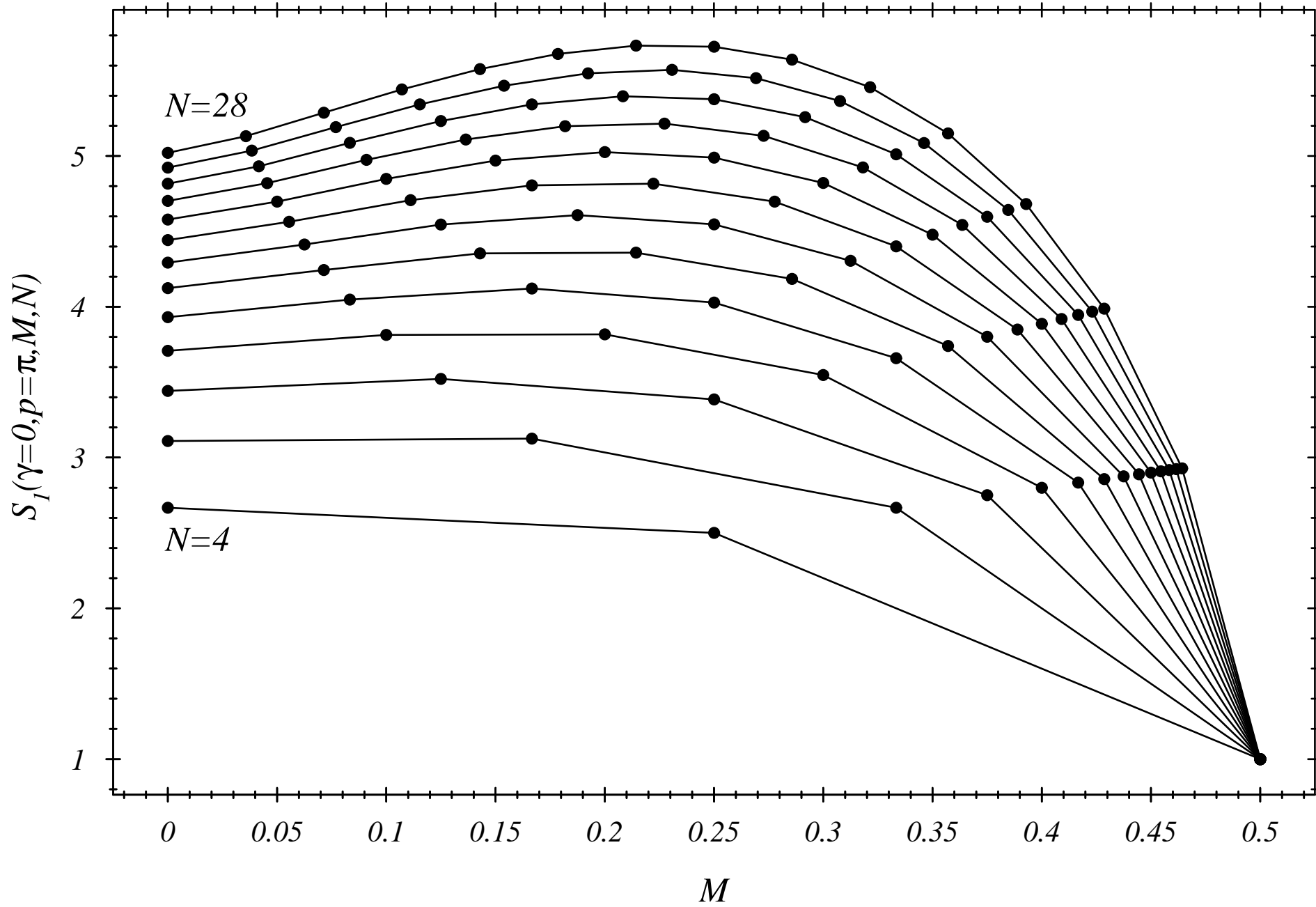
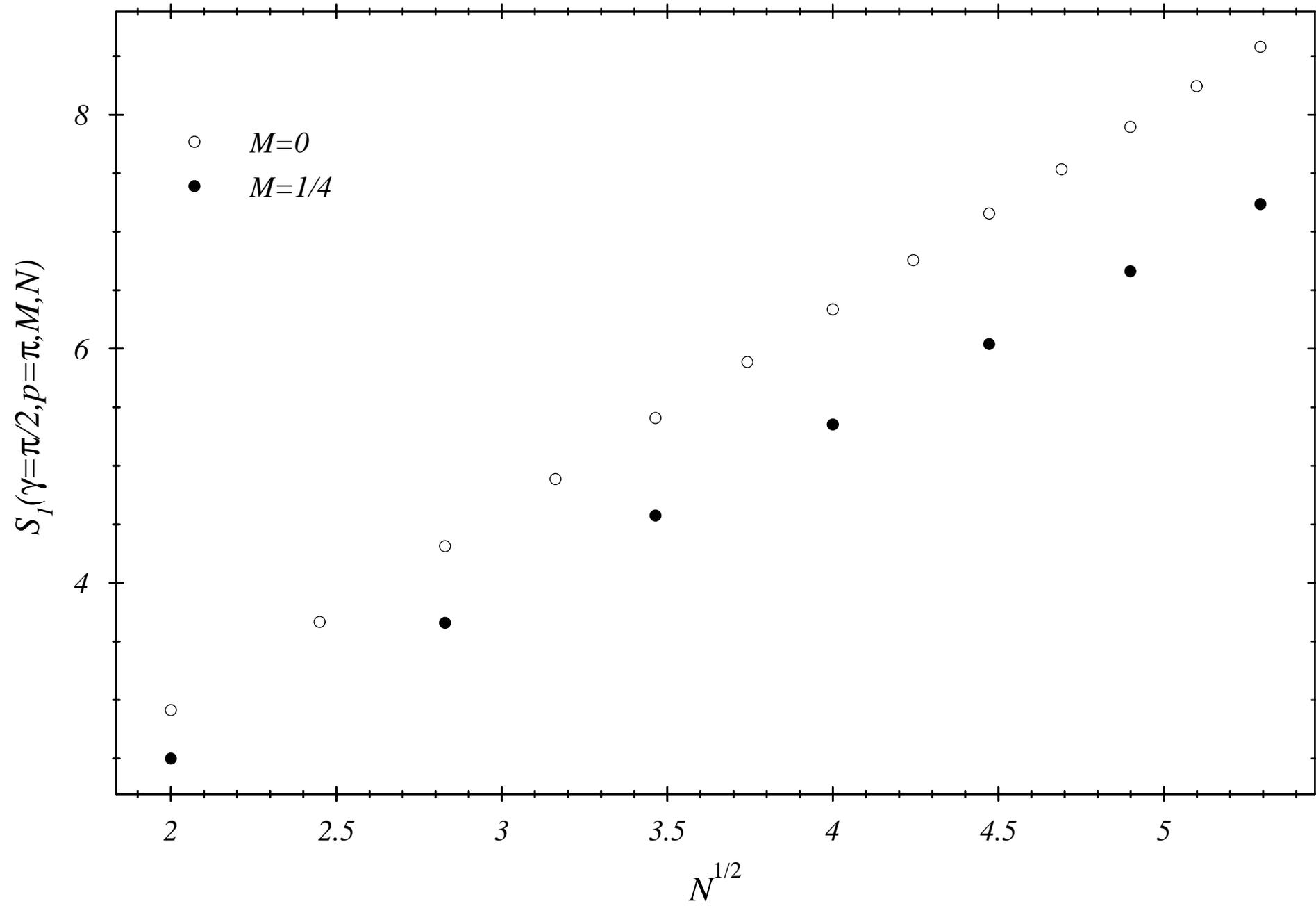
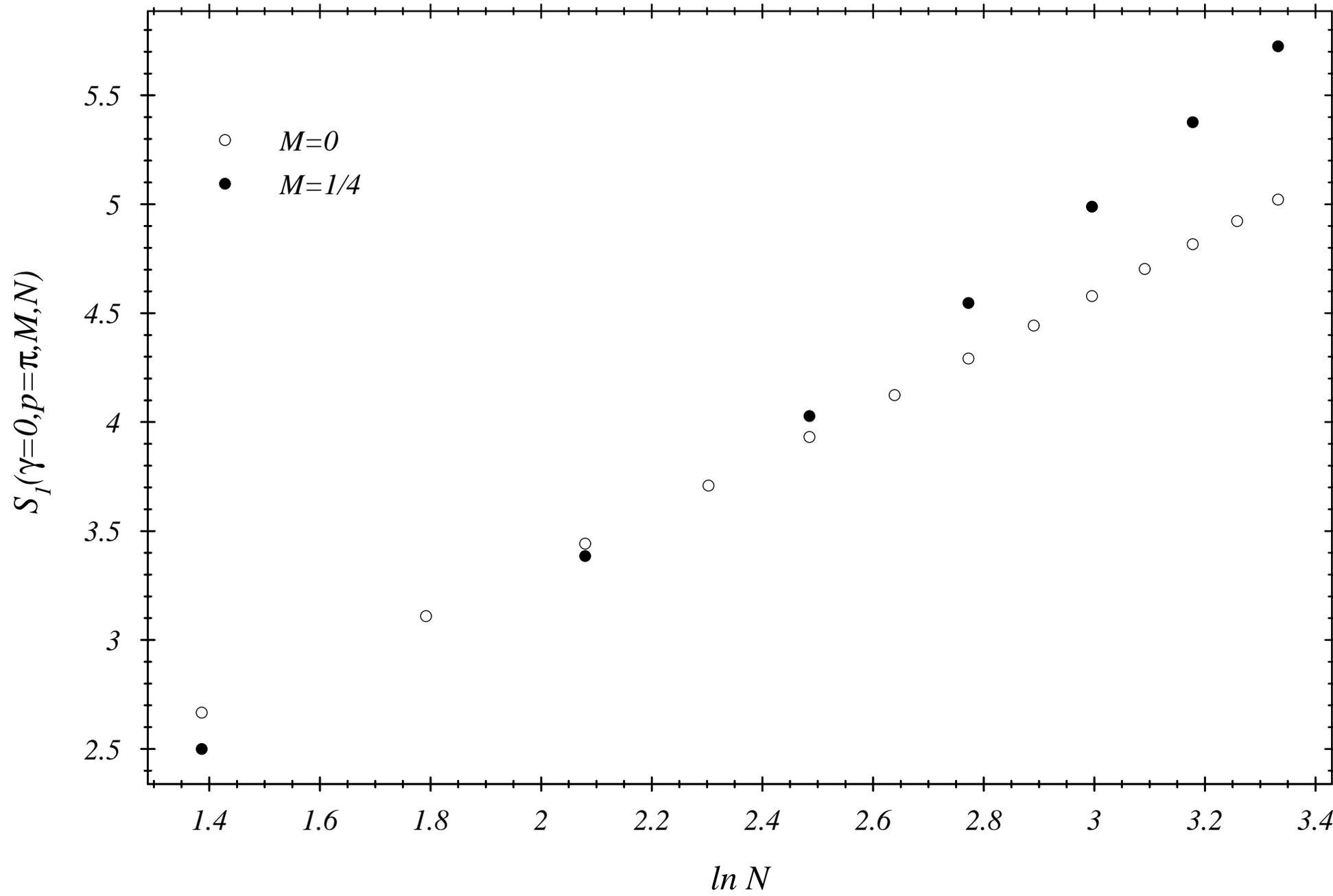
\end{document}